\documentclass[reprint,preprintnumbers,superscriptaddress,nofootinbib,showpacs,aps,prb,floatfix,longbibliography,onecolumn]{revtex4-2}
\usepackage{amsmath}
\usepackage{amssymb}
\usepackage{amstext}
\usepackage{amsopn}
\usepackage{amsfonts}
\usepackage{amsxtra}
\usepackage[pdftex]{color}
\usepackage{graphicx}
\usepackage{booktabs} 
\usepackage{xfrac}
\usepackage{multirow} 
\usepackage[table]{xcolor} 
\usepackage{makecell} 
\usepackage{longtable} 
\usepackage{float}
\usepackage{soul}
\usepackage{longtable}
\usepackage{dcolumn}
\usepackage{appendix}
\usepackage{bm}
\usepackage{url}
\usepackage{bibentry}
\usepackage[colorlinks=true, urlcolor=blue, linkcolor=blue, citecolor=blue, pdftex]{hyperref}
\PassOptionsToPackage{unicode}{hyperref}
\PassOptionsToPackage{naturalnames}{hyperref}

\usepackage[normalem]{ulem}

\usepackage{CJKutf8} 

\usepackage[mathlines]{lineno} 
\begin{document}
	
	\title{Electronic band structures of topological kagome materials}
	
	\author{Man~Li (\begin{CJK*}{UTF8}{gbsn}李满\end{CJK*})} 
	\affiliation{School of information network security, People's Public Security University of China, Beijing 100038, China}
	
	\author{Huan~Ma (\begin{CJK*}{UTF8}{gbsn}马欢\end{CJK*})} 
	\affiliation{Department of Physics, Key Laboratory of Quantum State Construction and Manipulation (Ministry of Education), and Beijing Key Laboratory of Opto-electronic Functional Materials $\&$ Micronano Devices, Renmin University of China, Beijing 100872, China}
	
	\author{Rui~Lou (\begin{CJK*}{UTF8}{gbsn}娄睿\end{CJK*})} \email{lourui09@gmail.com}
	\affiliation{Leibniz Institute for Solid State and Materials Research, IFW Dresden, 01069 Dresden, Germany}
	\affiliation{Helmholtz-Zentrum Berlin f\"{u}r Materialien und Energie, Albert-Einstein-Stra{\ss}e 15, 12489 Berlin, Germany}
	\affiliation{Joint Laboratory “Functional Quantum Materials” at BESSY II, 12489 Berlin, Germany}
	
	\author{Shancai~Wang (\begin{CJK*}{UTF8}{gbsn}王善才\end{CJK*})} \email{scw@ruc.edu.cn}
	\affiliation{Department of Physics, Key Laboratory of Quantum State Construction and Manipulation (Ministry of Education), and Beijing Key Laboratory of Opto-electronic Functional Materials $\&$ Micronano Devices, Renmin University of China, Beijing 100872, China}

	\date{\today}
	\begin{abstract}
		
		The kagome lattice has garnered significant attention due to its ability to host quantum spin Fermi liquid states. Recently, the combination of unique lattice geometry, electron-electron correlations, and adjustable magnetism in solid kagome materials has led to the discovery of numerous fascinating quantum properties. These include unconventional superconductivity, charge and spin density waves (CDW/SDW), pair density waves (PDW), and Chern insulator phases. These emergent states are closely associated with the distinctive characteristics of the kagome lattice's electronic structure, such as van Hove singularities, Dirac fermions, and flat bands, which can exhibit exotic quasi-particle excitations under different symmetries and magnetic conditions.
		Recently, various quantum kagome materials have been developed, typically consisting of kagome layers stacked along the $z$-axis with atoms either filling the geometric centers of the kagome lattice or embedded between the layers.
		In this topical review, we begin by introducing the fundamental properties of several kagome materials. To gain an in-depth understanding of the relationship between topology and correlation, we then discuss the complex phenomena observed in these systems. These include the simplest kagome metal $T_3X$, kagome intercalation metal $TX$, and the ternary compounds $AT_6X_6$ and $RT_3X_5$ ($A$ = Li, Mg, Ca, or rare earth; $T$ = V, Cr, Mn, Fe, Co, Ni; $X$ = Sn, Ge; $R$ = K, Rb, Cs). Finally, we provide a perspective on future experimental work in this field.
		
	\end{abstract}

	\maketitle


	
	
	\section{Introduction}

	The kagome lattice typically consists of a honeycomb border shared triangular lattice, thus the kagome physics inherits the honeycomb's Dirac and Van Hove singularity (vHS) features \cite{Liu_2014}. Considering the influence of spin-orbit coupling (SOC), the ideal kagome lattice is expected to yield an almost dispersionless, phase-destructive flat band (FB), exhibit markedly enhanced Coulomb interaction, and potentially manifest phenomena such as high-temperature ferromagnetism \cite{Tanaka2003,CYP2028}, Wigner crystal \cite{WCJ2007,Jaworowski_2018}, and the long-pursued high-temperature quantum anomalous Hall effect \cite{Tang2011,Xu2015}. Therefore, the exploration of a clean and ideal two-dimensional (2D) kagome material has received extensive attention in condensed matter physics in recent years.

\begin{figure*}[htb]
		\begin{center}
			\includegraphics[width=0.8\textwidth]{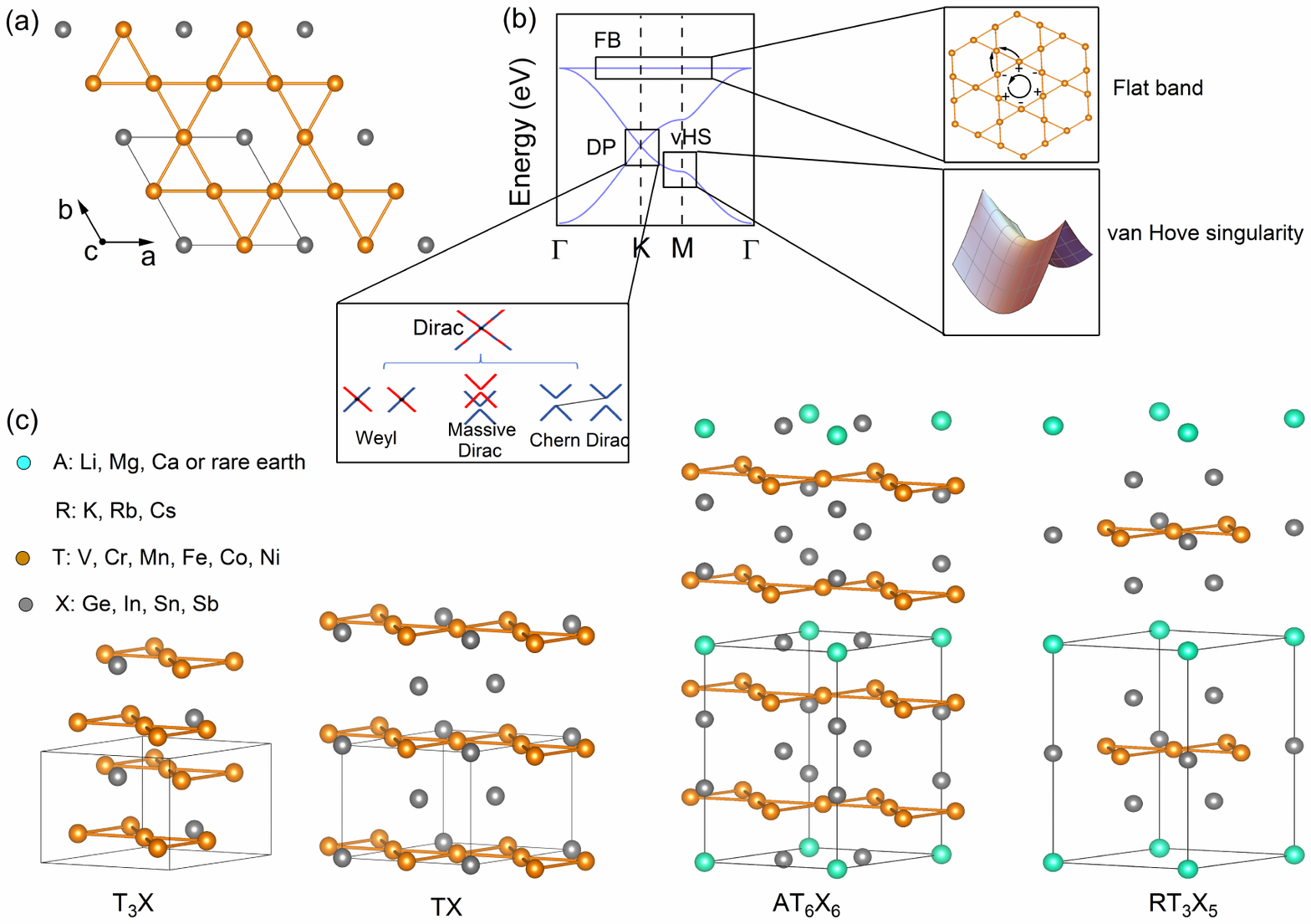}
			\caption{ 
				\textbf{Crystal structure and electronic structure of typical kagome metals}
					(a) Top view of the kagome plane with space-filling atoms (X: Ge, In, Sn, Sb) at the center of the hexagon. The in-plane unit cell is marked with the parallelogram.
					(b) Tight-binding band structure of the kagome lattice exhibiting a Dirac band at the K point, two van Hove singularities at the M point, and a flat band across the whole BZ. The illustrations represent the geometric phase-destructive flat band, saddle band, and the Dirac cone-derived Weyl/Dirac.
					(c) Stacking sequences of the kagome metal series $T_3X$, $TX$, $AT_6X_6$, and $RT_3X_5$, respectively.				
			}\label{F1}
		\end{center}
	\end{figure*}

	An ideal bulk kagome material consisting solely of kagome lattices does not exist. In practice, most of the topological kagome materials contain an additional atom in the center of the kagome lattice  (Fig.~\ref{F1}(a)), which could lead to additional hopping that modifies the ideal kagome band structure. Additionally, kagome solid materials can be constructed by selecting the category, method, and sequence of insertion layers between the kagome layer \cite{Mohamed2023}, where a pristine kagome layer can be achieved. Studying the clean kagome layer in this manner is primarily suited for surface-sensitive detection methods, such as angle-resolved photoemission spectroscopy (ARPES) and scanning tunneling microscopy/spectroscopy (STM/STS).

	By incorporating extra atomic layers between the kagome interlayer, the material's overall properties become more intricate, potentially giving rise to more complex strongly correlated phases. Unlike the topological systems based on s or p orbitals near the non-interacting limit, in these intermetallic materials, the kagome lattice filled with the low-energy 3d electrons of transition metals could provide diverse magnetic, orbital, and electronic properties \cite{Jiang2022, Wang2023, YJX2022}. In light of this, leveraging diverse symmetries, magnetism, and electronic correlation characteristics could offer a promising avenue for investigating the interplay between topological and strong correlation effects in bulk kagome materials. For example, the kagome magnetic Weyl fermions resulting from breaking the time-reversal symmetry \cite{Wang2018,LDF2019}, the Chern phase manifesting in ferromagnetism \cite{Xu2015}, and the strongly entangled CDW and superconducting phases associated with the correlation effects of the vHSs\cite{Yu2012,Kiesel2013,Feng2021} and/or the FBs\cite{LYD2024,Guo2024}, as seen in Fig.~\ref{F1}(b).
	
	Recently, numerous transition metal kagome materials across various families have emerged, leading to extensive studies and fruitful results. 
	In general, compounds within the same family display homologous properties due to their shared structure and similar magnetism, while those belonging to different series possess distinct structures, resulting in divergent properties. 
	For example, emergent superconductivity, CDW/SDW/PDW phases, nematic and stripe orders are observed in the topological kagome metals $R$V$_3$Sb$_5$ ($R$ = K, Rb, Cs) family \cite{Jiang2022}. The intrinsic anomalous Hall effect (AHE) and nontrivial topological band structure are shown in most of the $A$Mn$_6$Sn$_6$ family of compounds \cite{LLF2024}, and the $\mathbb{Z}_2$ topological state presents in the $A$V$_6$Sn$_6$ family \cite{Wang2023}. Meanwhile, similar properties are also shared among different kagome families, implying a common physical origin and/or quasiparticle excitation. 
	Examples include the magnetic Weyl states in Mn$_3$Sn/Mn$_3$Ge and Co$_3$Sn$_2$S$_2$ deriving from the Dirac cone of the kagome lattice \cite{Kuroda2017,LDF2019}, the negative orbital magnetism in Co$_3$Sn$_2$S$_2$ and CoSn compounds deriving from the FB of the kagome lattice \cite{Yin2019,Huang2022}, and the CDW states in $R$V$_3$Sb$_5$ and FeGe compounds deriving from the vHSs of the kagome lattice \cite{Wang2023}.
	Thus, performing comparative analyses of kagome compounds across different family systems to discern similarities and differences in physical properties will deepen our understanding of the topology and electron correlation effects within the kagome platform.
	
	In this review, we will focus on the recent advancements in understanding the electronic structure, magnetism, CDW instabilities, and superconductivity in several prototypical kagome materials. We start with the $T_3X$ family of compounds in Section \ref{T3X}, where we emphatically review the tunability of the kagome band structure via spin reorientation in ferromagnetic Fe$_3$Ge. Next, we review the $TX$ kagome family represented by FeSn, CoSn, and FeGe compounds, discussing their diverse magnetic and band features in Section \ref{TX}. Then we examine the electron correlations and topological properties in $AT_6$Sn$_6$ in Section \ref{AT166}. In Section \ref{Origins}, we explore the underlying origins of different exotic phases, such as CDW, SDW, PDW, and unconventional superconductivity, as well as their potential entanglement in the $R$V$_3$Sb$_5$ family. Finally, we conclude the review by providing a future perspective on the research field of kagome lattices. The detailed properties of these kagome materials are summarized in Table~\ref{kagome_properties}.

	\section{The simplest kagome lattice in $T_{3}X$ family} \label{T3X}
	
	The $T_3X$ ($T$ = Mn, Fe, Ni; $X$ = Ge, Sn, In) family hosts the simplest kagome structure, characterized by a lattice that is stacked exclusively along the $z$-axis, with the Sn/Ge/In atom filling the center of the hexagon. 
	Despite the strong interlayer interaction between the kagome layers, the properties of these solid materials are predominantly influenced by kagome features. 
	These features include a Dirac-like band connected to the saddle point, capped by a FB.
	The Dirac states in the kagome lattice can split into two Weyl points with opposite chirality upon the breaking of time-reversal symmetry. This results in the formation of Weyl semimetals, whose surface states are characterized by Fermi arcs that connect the two Weyl points.
	
	\begin{figure*}[htb]
		\begin{center}
			\includegraphics[width=0.75\textwidth]{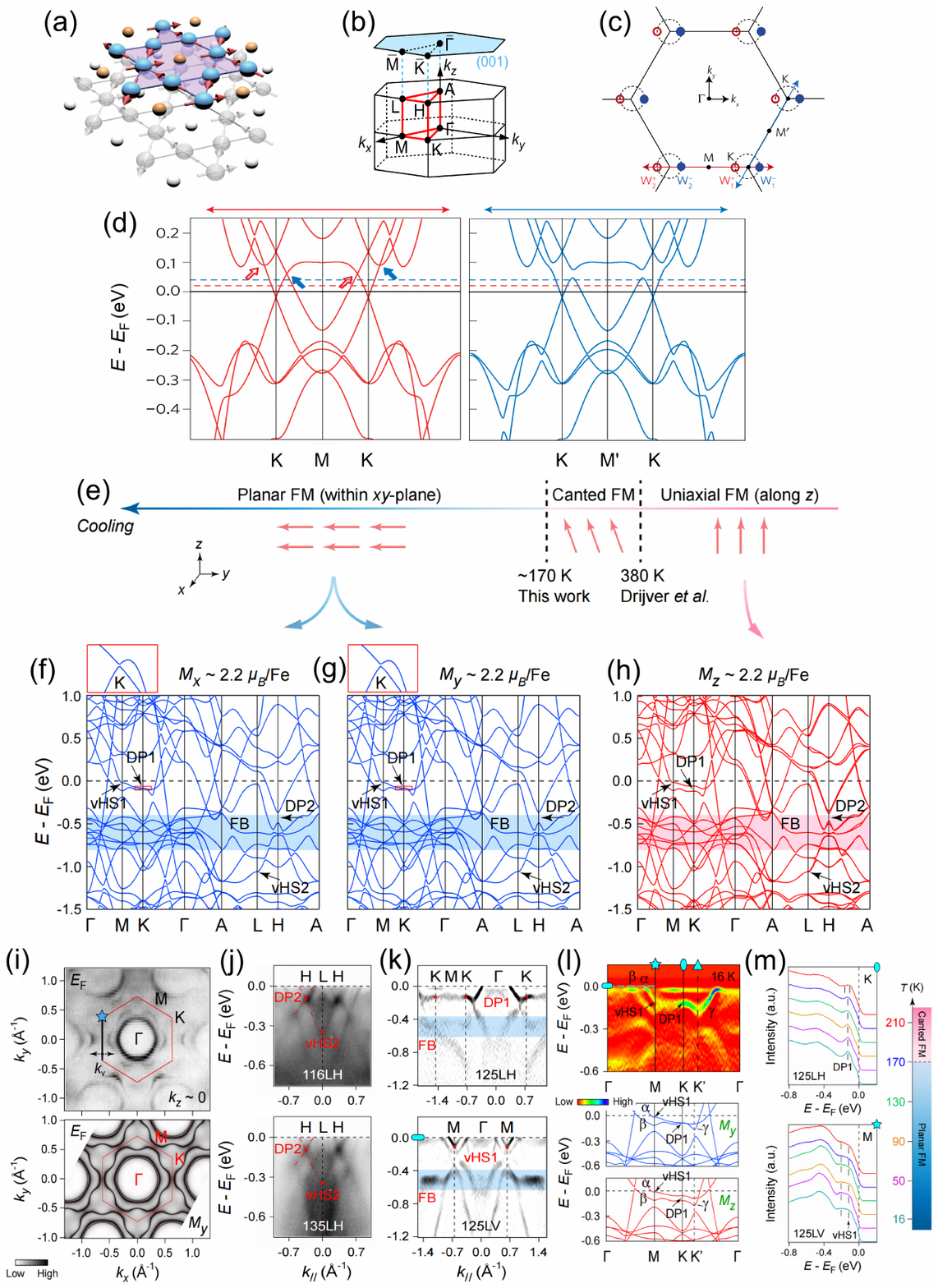}
			\caption{ \textbf{Lattice structure and electronic band structure of $T_3X$.}
				(a) Magnetic texture of Mn$_3$Sn with a 120$^\circ$ structure of Mn moments lying in each $x$–$y$ kagome plane.
				(b) 3D and projected BZs of $T_3X$ with marked high-symmetry points.
				(c) Distribution of the Weyl points in the bands on the $k_x$–$k_y$ plane at $k_z$ = 0 near $E_{\mathrm{F}}$ for the magnetic texture shown in Fig.~\ref{F2}a.
				(d) DFT calculated the band structure of the Weyl point along the high-symmetry line $K$--$M$--$K$ in Mn$_3$Sn (Part a, c and d adapted from Ref.\cite{Kuroda2017}).
				(e) A summary of the magnetic phase transitions as a function of temperature in Fe$_3$Ge.
				(f), (g), (h) DFT calculated band structures of Fe$_3$Ge in the FM phase for the Fe moments ($\mu_{\mathrm{Fe}}$ $\sim$ 2.2 $\mu_{\mathrm{B}}$) aligned along the $x$, $y$, and $z$-axes, respectively.
				(i) FS mapping of Fe$_3$Ge taken by the 125-eV photons ($k_z$ = 0 plane) with linear horizontal (LH) polarization and DFT calculated bulk FSs in the $k_z$ = 0 plane with an FM moment aligned along the $y$-axis, respectively.
				(j) ARPES intensity plots measured along the $H$--$L$--$H$ directions with photon energies of 116 and 135 eV, respectively (LH polarization). The red dashed lines are guides to the eyes for the Dirac cone structure of the DP2 at the $H$ point and its connection to the vHS2 at the $L$ point.
				(k) Second derivative intensity plots taken along the $K$--$\Gamma$--$K$ and $M$--$\Gamma$--$M$ directions with LH and LV polarizations, respectively. The less-dispersive DP1 at the $K$ point, the vHS1 at the $M$ point, and the FB regions are marked out.
				(l) Second derivative intensity plots of Fe$_3$Ge along the $\Gamma$--$M$--$K$--$\Gamma$ lines with $h\nu$ = 125 eV and T = 16 K (upper panel). Comparison between the calculated bands near $E_{\mathrm{F}}$ of FM Fe$_3$Ge with the Fe moments ($\mu_{\mathrm{Fe}}$ $\sim$ 2.2 $\mu_{\mathrm{B}}$) along the y and z axes (lower panel).
				(m) Temperature-dependent EDCs taken at $K$ and $M$ points with LH and LV polarizations as indicated in Fig.~\ref{F2}l, respectively \cite{Lou2023}.
			}\label{F2}
		\end{center}
	\end{figure*}
	
	Mn$_3$Sn and Mn$_3$Ge exhibit a non-collinear antiferromagnetic (AFM) order transition at approximately T = 420 K (Fig.~\ref{F2}(a)). Upon cooling below this temperature, the Mn magnetic moments form an inverse triangular spin configuration, resulting in a nonzero Berry curvature and producing a weak net magnetization with small magnetic anisotropy within the kagome plane \cite{Kuroda2017,Higo2022,Kimata2019}. Meanwhile, the chiral antiferromagnets Mn$_3$Sn and Mn$_3$Ge have been found to exhibit a surprisingly large room temperature AHE \cite{Nakatsuji2015}. 
	
	Recent theoretical and experimental advances have demonstrated that the intrinsic AHE in Mn$_3$Sn and Mn$_3$Ge is driven not by magnetism but by the Berry curvature of the nontrivial topological Weyl states \cite{RMP20101}. Benefiting from ARPES measurements and theoretical efforts, the AFM kagome compounds Mn$_3$Sn and Mn$_3$Ge have been shown to host time-reversal symmetry-broken Weyl fermions, as illustrated in Fig.~\ref{F2}(c) and \ref{F2}(d) \cite{Kuroda2017,Yang2017}. 
	
	These Weyl states with diverging Berry curvatures support distinct anomalous transport phenomena, including the chiral anomaly effect, nonlocal transport, Weyl orbit transport, negative longitudinal magnetoresistance, anomalous Nernst effect, thermal Hall effect, magneto-optical Kerr effect, and both topological and planar Hall effects \cite{Nielsen1983,Nandy2017,West1963,PHE2018,mao2019,RMP2018,Ikhlas2017,LXK2017,Higo2018,Rout2019,XLC2020,Burkov2017}. 
	
	Furthermore, measurements of the scanning tunneling spectroscopy (STS) spectrum in Mn$_3$Sn reveal a many-body resonance akin to the Kondo effect, stemming from the robust Coulomb interaction between the phase-destructive FB and the itinerant conduction band near $E\rm_F$. These results suggest that Mn$_3$Sn is a correlated topological kagome metal with a flat band state \cite{ZSTS2020}. 
	
	Given that Mn$_3$Sn samples can possess different magnetic structures depending on their chemical compositions and/or growth conditions, an external magnetic field can effectively adjust the intrinsic AHE and the electronic band structure. Mn$_3$Sn thus provides a promising platform for establishing the correlations between space-time spin texture, AHE, and band topology \cite{Sung2018,Li2023}.

	Another compound, Fe$_3$Ge, is isostructural to Mn$_3$Sn and Mn$_3$Ge, crystallizing in a hexagonal structure with the $P\rm{6_3}/mmc$ (No. 194) space group and belonging to the kagome lattice binary $T_3X$ series. In contrast to the antiferromagnet Mn$_3$Sn and Mn$_3$Ge, the kagome ferromagnet Fe$_3$Ge does not exhibit Weyl states but demonstrates rich characteristics in the temperature evolution of spin texture and topology \cite{YU2022,Lou2023,Drijver1976,ZZX2024}. Due to the strong interplane coupling in the $T_3X$ series, obtaining atomically flat surfaces by cleaving the single crystals for surface-sensitive experiments is almost unfeasible. It is usually necessary to polish the (001) surfaces and then sputter and anneal the polished surfaces in a vacuum.
	
	As depicted in Fig.~\ref{F2}(e), it was reported that the Fe moments align along the z-axis at high temperatures and then start to cant away from the $z$-axis towards the $xy$-plane at a lower temperature (T $\sim$ 380 K), finally forming a planar ferromagnetic (FM) ground state at T $<$ 170 K \cite{Lou2023}. 
	With increasing temperature in Fe$_3$Ge, spins tend to align with the easy axis, which facilitates the formation of a non-coplanar spin texture under an external magnetic field, leading to the emergence of a time-space topological Hall effect (THE) resembling that of Mn$_3$Sn \cite{ZZX2024,Rout2019}. 
	
	Fe$_3$Ge exhibits complex magnetic phase transitions, thus providing a good platform for studying the correspondence between the intricate effects of magnetic spin configurations and electronic structure. In Figs.~\ref{F2}(f)-\ref{F2}(g), the overall bulk band structure in the FM state is calculated from density functional theory (DFT), with the preferred spin direction along the $x$, $y$, and $z$ axes, respectively. It exhibits an in-plane ferromagnetic state at low temperatures and shows minimal discrepancy in the electronic structure when the Fe moment's easy plane aligns along the $x$ or $y$-axis, suggesting they are both ground states in calculations.
	
	Meanwhile, as the Fe moments cant from the $z$-axis into the $xy$-plane upon cooling, the two types of kagome-derived Dirac fermions respond differently. The Dirac point (DP1) with less dispersive bands in the $k_z \sim 0$ plane, containing the 3$d_{z^2}$ orbitals, evolves from gapped to nearly gapless. In contrast, the other Dirac point (DP2) with linear dispersions in the $k_z \sim \pi$ plane, which embraces the 3$d_{xz}$/3$d_{yz}$ components, remains intact, indicating that the effect of spin reorientation on the Dirac fermions is orbital selective \cite{Lou2023}. Comparing the DFT calculations with ARPES, the experimental data and theoretical calculations exhibit excellent consistency in Figs.~\ref{F2}(i)-\ref{F2}(l). 
	
	As shown in Fig.~\ref{F2}(m), the temperature evolution of energy distribution curves (EDCs) taken at the $K$ and $M$ points demonstrates that the gapless DP1 at the $K$ point in the planar FM state becomes gapped as the spin cants towards the $z$-axis, while the vHS at the $M$ point remains almost unchanged. This remarkable change in the DP1 gap, primarily involving the 3$d_{z^2}$ orbitals, resembles that of the Kane-Mele type SOC gap \cite{Kane2005}, which is negligible under in-plane magnetic order but strongly enhanced under out-of-plane magnetic order \cite{Lou2023}. It is thus expected that the spontaneous magnetization of Fe$_3$Ge could induce a Chern Dirac state at high temperatures. This behavior also mirrors the strong coupling of Dirac fermions with the $z$-axis vector field observed in Fe$_3$Sn$_2$ \cite{Yin2018}.
	
	Moreover, comparative studies between the non-charge-ordered Fe$_3$Ge and its sibling compound charge-ordered FeGe strongly support that the orbital-selective vHSs (3$d_{xz}$/3$d_{yz}$) near the Fermi level play an indispensable role in driving the novel charge order on a magnetic kagome lattice, as in FeGe \cite{Teng2023}. These unambiguous observations provide a feasible route to design and manipulate the mass of Dirac fermions and the vHSs for realizing exotic quantum phases in the kagome magnet Fe$_3$Ge and its siblings.

	\begin{table*}[!htbp] 
		\caption{Magnetic properties, band topology, and intrinsic AHE of various categories within the kagome family. }
		\label{kagome_properties} 
		\centering
		\begin{tabular}{llll}
			\toprule
			\midrule
			Material & \phantom{~~~~~~} Low-temperature magnetic order & \phantom{~~~~~~} THE and intrinsic AHE & \phantom{~~~~~~} Band features   \\
			\midrule
			Mn$_3$Sn/Ge & \phantom{~~~~~~} Chiral AFM\cite{Kuroda2017,Kimata2019} & \phantom{~~~~~~} THE/AHE\cite{Rout2019,Nakatsuji2015} & \phantom{~~~~~~}	Weyl semimetal, flat band\cite{Kuroda2017,ZSTS2020};
			\\
			Fe$_3$Ge & \phantom{~~~~~~} planar FM \cite{Drijver1976} &	\phantom{~~~~~~} THE\cite{ZZX2024} & \phantom{~~~~~~} Dirac cone, flat band and vHS\cite{Lou2023}; 
			\\
			\midrule
			Fe$_3$Sn$_2$ & \phantom{~~~~~~} Non-collinear FM \cite{WQ2016} & \phantom{~~~~~~} THE/AHE\cite{WQ2016,DQH2022} &	\phantom{~~~~~~} Massive Dirac and flat band\cite{Ye2018,LZY2018};
	    	\\
	    	\midrule
	    	FeSn & \phantom{~~~~} AFM\cite{Sales2019} & \phantom{~~~~~~} Not reported & \phantom{~~~~~~} Dirac cone, flat band\cite{Kang2019,LZY2020};
	    	\\
	    	FeGe & \phantom{~~~~} AFM\cite{Teng2022} & \phantom{~~~~~~} AHE\cite{Teng2022} & \phantom{~~~~~~} Dirac cone, vHS and CDW\cite{Teng2022,Teng2023};
	    	\\
	    	CoSn & \phantom{~~~~} PM\cite{Xie2021} & \phantom{~~~~~~} Not reported & \phantom{~~~~~~} Dirac cone and flat band\cite{Liu2020,Kang2020,Huang2022}
	    	\\
	    	\midrule
			$A$Mn$_6$Sn$_6$
			& \phantom{~~~~} 
			\makecell[l]{FM:$A$ = Li, Mg, Ca, Nd, Sm\\ and Yb\cite{Mazet2006,Chen2021,Ma2021,Mazet2002,Song2024,LLF2024};\\
			(Helical) AFM :$A$ = Sc, Y, Lu\cite{Chafik1991,ZH2022,WQ2021,Ghimire2020,MWL2021};\\
			FIM: $A$ = Tb, Dy, Ho, Er, Tm \\ and Gd\cite{Gao2021,Venturini1991,Xu2022,Zeng2022,Kabir2022,Dhakal2021,Suga2006,Riberolles2024,Wang2022,Gorbunov2012};} 
			&  \phantom{~~~~~~}
			\makecell[l]{THE in $A$ = Y, Er, Sc and \\ Tm \cite{Ghimire2020,WQ2021,Dhakal2021,ZH2022,Liu2023};\\ AHE except for LuMn$_6$Sn$_6$\\ \cite{Gao2021,Asaba2020,WQ2021,MWL2021,Dhakal2021,Chen2021,Ma2021,Zeng2022,ZH2022,Lv2023,LLF2024};\\}
			& \phantom{~~~~~~} 
			\makecell[l]{Dirac cone and flat band in \\$A$ = Y, Gd, Tb, Dy\cite{Li2021,LZH2021,Gu2022}; \\ vHS in YMn$_6$Sn$_6$\cite{Li2021};\\ Chern Dirac in TbMn$_6$Sn$_6$\cite{YGX2020};}
			\\
			\midrule
			$A$V$_6$Sn$_6$
			& \phantom{~~~~~~} 
		    \makecell[l]{PM: $A$ = Y, Lu\cite{Pokharel2021,Lee2022};\\
		    No magnetic ordering in $A$= Er, Tm \\and Sm
		    at low temperature\cite{Pokharel2021,Lee2022,Huang2023}; \\Weak magnetism in $A$ = Gd, Tb, Dy \\and Ho \cite{Lee2022,Pokharel2022,Ishikawa2021,Pokharel2022,ZXX2022};}
			&  \phantom{~~~~~~}
			AHE reported in ScV$_6$Sn$_6$\cite{Mozaffari2024}
			&  \phantom{~~~~~~}
			\makecell[l]{Dirac cone and vHS in $A$ = Gd,\\ Ho, Tb and Sc\cite{Peng2021,Sante2023,Hu2022,Hu2024};\\ Flat band in Gd and Tb\cite{Peng2021,Sante2023};\\ Mostly $\mathbb{Z}_2$ surface state\cite{Pokharel2021,Hu2022,Sante2023,Peng2021,Hu2024};\\ CDW in ScV$_6$Sn$_6$\cite{Arachchige2022}; } 
			\\
			\midrule
			$R$V$_3$Sb$_5$
			&  \phantom{~~~~~~} \makecell[l]{No long-range order: \\ $R$ = K, Rb and Cs\cite{Mielke_2022,Khasanov2022};}
			&  \phantom{~~~~~~} AHE\cite{Yang2020,ZXB2022}
			&  \phantom{~~~~~~} \makecell[l]{Dirac cone and vHS\cite{Ortiz_202102,LZH_2021,2022Charge-LouRui};\\ CDW, 
			PDW, nematic/stripe orders \\ and superconducting \cite{Neupert2022,Wang2023};}
			\\
			\midrule
			(Rb/Cs)V$_8$Sb$_{12}$
			& \phantom{~~~~~~} PM \cite{YYX_2021,YQW_2021}
			& \phantom{~~~~~~} Not reported
			& \phantom{~~~~~~} \makecell[l]{Dirac nodal line\cite{Cai2023};\\ vHS \cite{Cai2023,2023Mahuan-CsV8Sb12PRB};}
			\\
			\midrule
			(Rb/Cs)V$_6$Sb$_{6}$
			& \phantom{~~~~~~} PM \cite{YYX_2021,YQW_2021}
			& \phantom{~~~~~~} Not reported
			& \phantom{~~~~~~} \makecell[l]{Dirac nodal line and pressure-\\ induced superconductivity\cite{SMZ_2021};}
			\\
			\midrule
			Co$_3$Sn$_2$S$_2$
			& \phantom{~~~~~~}
			FM\cite{LEK2018}
			& \phantom{~~~~~~}
			AHE\cite{Wang2018}
			& \phantom{~~~~~~}\makecell[l]{Wely semimetal\cite{LDF2019} and flat band\cite{Yin2019}}
			\\
			Ni$_3$In$_2$S$_2$
			& \phantom{~~~~~~}
			PM\cite{ZTT2022}
			& \phantom{~~~~~~}
			Not reported
			& \phantom{~~~~~~}
			Dirac nodal line\cite{ZTT2022};
			\\
			Ni$_3$In$_2$Se$_2$
			& \phantom{~~~~~~}
			AFM \cite{CL2023}
			& \phantom{~~~~~~}
			Not reported
			& \phantom{~~~~~~}
			Dirac nodal line\cite{Pradhan2024};
		    \\
			\midrule
			\bottomrule
		\end{tabular}
	\end{table*}

	\section{Diversified properties in kagome metals $TX$ system} \label{TX}

	In $T_3X$ serial compounds, such as Mn$_3$Sn and Fe$_3$Ge, strong three-dimensional properties and small momentum-size flat bands are revealed. These features primarily originate from the strong interlayer coupling between the kagome layers \cite{ZSTS2020,Lou2023}. 
	The strategic construction of low-dimensional electronic structures involves intercalating atomic layers, which can reduce the coupling interactions between the kagome layers. This may result in more quasi-two-dimensional kagome properties, larger momentum scales, and flat bands with finite bandwidth. For example, quasi-two-dimensional Dirac-like band structures have been observed in kagome lattice Fe$_3$Sn$_2$ (denoted as the "32" system) \cite{Ye2018,Ye2019}, which can be viewed as inserting a layer of Sn atoms between two neighboring Fe$_3$Sn kagome layers. Fe$_3$Sn$_2$ features a non-collinear ferromagnetic configuration, with competing magnetic interactions and magnetic anisotropy that preserve a robust THE \cite{DQH2022}. The intrinsic AHE signals topologically non-trivial properties\cite{WQ2016}, and the massive Dirac gap can be further tuned by the applied magnetic field, offering a model system for exploring topological phases related to the Kane–Mele term \cite{Yin2018}. The phase-destructive FB has slao been reported in Fe$_3$Sn$_2$\cite{LZY2018}. The "32" system, serving as an intermediate structure between the $T_3X$ and $TX$ systems, can be converted into the $TX$ system by inserting Sn/Ge atoms between each nearest-neighboring double kagome layers.
	
	In the $TX$ system, a chemical stoichiometry ratio of 1:1 is achieved by inserting a hexagonal layer of Sn atoms into the $T_3X$ basic framework.
	Furthermore, compared to the $T_3X$ serial compounds, the structural characteristics of $TX$ typically lead to two possible surface terminations when cleaved in situ: the kagome ($T_3X$) termination and the Sn/Ge termination, as illustrated in Fig.~\ref{F1}(c).
	In this section, we mainly introduce three compounds of the $TX$ series: FeSn, FeGe, and CoSn
	
	\begin{figure*}[htb]
		\begin{center}
			\includegraphics[width=1\textwidth]{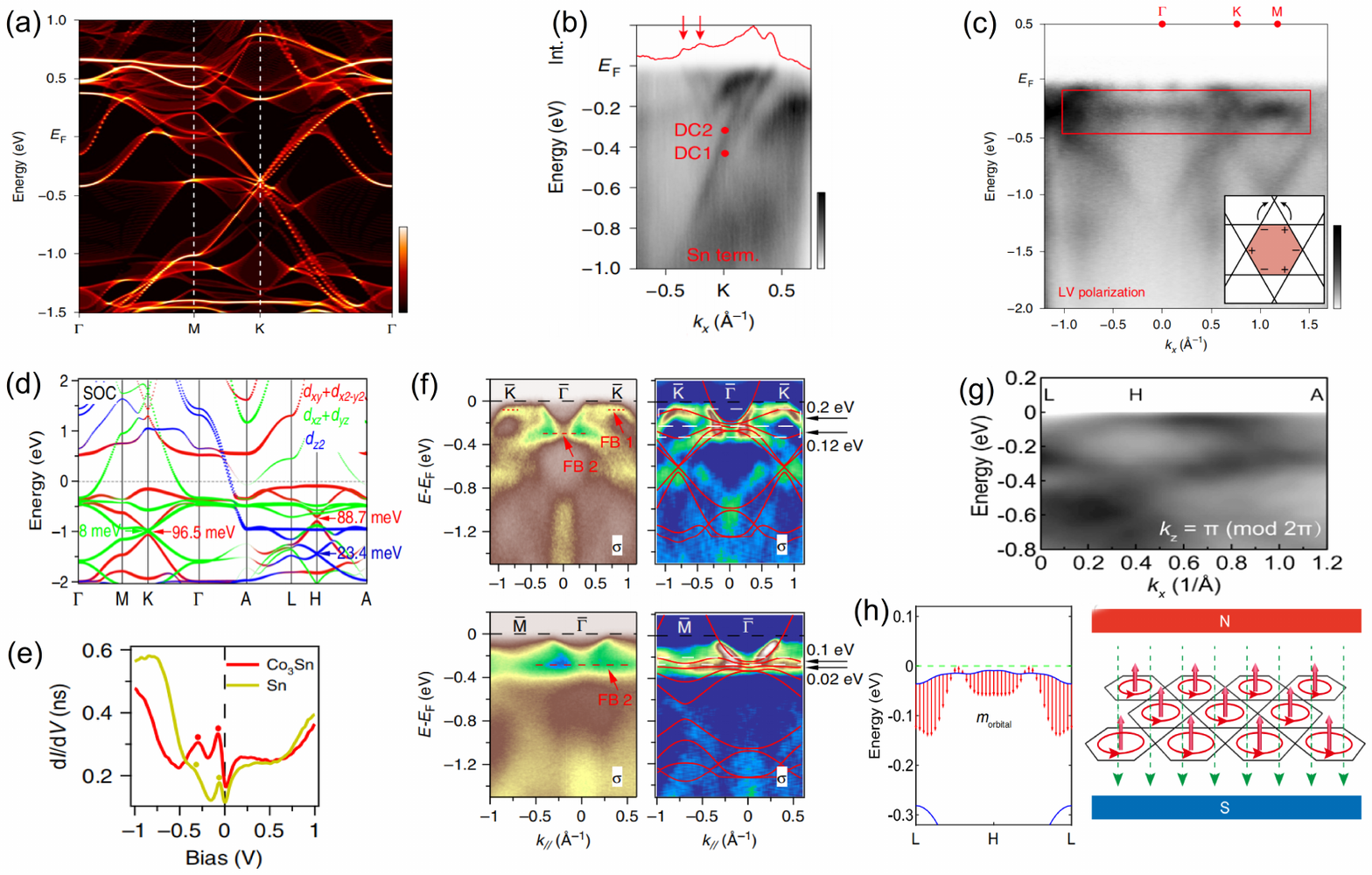}
			\caption{ \textbf{Crystal structure and electronic structure of $TX$.}
				(a) Energy-momentum dispersion of FeSn integrated along the $k_z$ direction.
				(b) Energy-momentum dispersion of FeSn across the $K$ point on Sn termination. DC1 and DC2 indicate the position of Dirac points.
				(c) Experimental band structure of FeSn along the $\Gamma$--$K$--$M$ high-symmetry direction measured with LV polarization on kagome termination. The red box highlights the nearly flat band with a binding energy of $\sim$ 0.23 eV (Part a-c adapted from Ref.\cite{Kang2019}).
				(d) DFT calculated bands of CoSn along high-symmetry directions with SOC. Orbital characters are indicated by the different colors. Flat bands, Dirac points, and band gaps induced by SOC at the Dirac points are indicated.
				(e) Scanning tunneling spectra taken on the Co$_3$Sn surface (red) and the Sn surface (yellow) ($V$ = 1 V, $I$ = 100 pA).
				(f) Intensity plots and corresponding second derivative plots along the $\Gamma$--$K$ and $\Gamma$--$M$ directions in $\sigma$ geometry, respectively. Flat bands are indicated by the red dashed lines (left panel). The red lines on the corresponding second derivative plots are DFT-calculated bands renormalized by a factor of 1.5 \cite{Liu2020}.
				(g) Highly resolved ARPES results along the $L$--$H$--$A$ directions with photon energies of 45 eV, showing a flat band with a high density of states around $E_{\mathrm{F}}$.
				(h) Calculated orbital moment of the flat band along the $L$--$H$--$L$ direction, with red arrows indicating the direction and magnitude of the magnetic moment (left). Schematic diagram of the magnetic-field-induced diamagnetism in the kagome lattice (right)(Part g-i adapted from Ref.\cite{Huang2022}).
			}\label{F3}
		\end{center}
	\end{figure*}

	FeSn exhibits magnetic ordering, with ferromagnetically aligned Fe moments within each kagome plane that are antiferromagnetically coupled along the $z$-axis \cite{Hartmann_1987,Ku1981,LZY2020,Kang2019}. 
	This configuration breaks both $P$ and $T$ symmetries individually while preserving the combined $PT$ symmetry. When combined with the nonsymmorphic symmetry $S_{2z}$, it provides an ideal platform to realize the symmetry protection of massless Dirac fermions even in the presence of SOC \cite{LZY2020}.
	
	Furthermore, it has been demonstrated that the spin degree of freedom in the current AFM kagome system can effectively serve as a mechanism to tune topological behaviors via doping and temperature, enabling the realization of massive/massless Weyl or Dirac states \cite{LZY2020,Moore2022,Meier2019,XT2022}. The breaking of the combined $PT$ symmetry on the surface results in a massive surface Weyl band \cite{Moore2022}.
	
	The tunable symmetry-breaking magnetic moment, $S_{2z}$, allows for the manipulation of the bulk Dirac gap and energy shift through doping and temperature adjustments. This tunability has been confirmed in cobalt-doped FeSn (Fe$_{1-x}$Co$_x$Sn) \cite{Moore2022,Meier2019,XT2022}. Figure~\ref{F3}(a) represents the DFT-calculated bulk bands integrated along $k_{z}$, revealing a distinct FB at about 0.5 eV above $E\rm_F$ and a gapless bulk Dirac Fermion at the $K$ point.
	
	ARPES measurements yielded two distinct terminations, namely the kagome and the Sn honeycomb, respectively. In the former case, experimental observations revealed a bulk Dirac state at the $K$ point formed by the unique geometric structure of kagome. In the latter scenario, not only was a kagome-derived bulk Dirac state measured, but also a surface Dirac state, as a surface resonant state of bulk Dirac, formed by the topmost ferromagnetic kagome layer under the surface potential, as shown in Fig.~\ref{F3}(b)\cite{Kang2019}.
	
	Orbital analysis reveals that the surface and bulk Dirac states both have an identical orbital character, $d_{xy}$+$d_{x^2-y^2}$. Photon energy-dependent ARPES measurements further reveal that all Dirac states reflect the 2D nature in FeSn. Additionally, the kagome-derived FB has also been detected by linear vertical (LV) polarization, as shown in Fig.~\ref{F3}(c). The spectral weight is uniformly distributed across almost the entire 2D Brillouin zone (BZ), except around the $K$ points \cite{Kang2019}.
	
	Based on the discussion, FeSn might be an ideal system where the Dirac Fermion and FB from in-plane $d$ orbitals are invulnerable to inter-layer interactions and retain their dispersionless character in momentum space. Intercalated atomic layers can reduce the coupling interaction between kagome layers to form an ideal 2D kagome structure. Thus, the Fe$_{1-x}$Co$_x$Sn system provides an ideal opportunity to tune the magnetic state while observing the interplay between magnetism and topologically protected electronic states.

FeGe, a kagome magnet with a CDW, is a sibling compound of FeSn and has recently attracted significant attention. The kagome lattice in FeGe exhibits collinear A-type  AFM order with FM moments in each layer parallel to the $z$-axis below $T_N$ $\sim$ 410 K. 
It transitions to a $z$-axis canted AFM structure below T = 60 K, while an exotic short-range CDW emerges within the AFM phase below T = 100 K \cite{Bernhard_1984,Bernhard_1988,Teng2022}. 
The CDW in FeGe enhances the moment of the collinear AFM order and induces an  AHE with a magnitude similar to that found in $R$V$_3$Sb$_5$ \cite{Yang2020,Yu2021}, potentially consistent with a chiral flux phase of circulating currents \cite{Teng2022}.

In FeGe, electronic correlations within the kagome-derived FB near the Fermi level in the paramagnetic state could drive the system into antiferromagnetically coupled FM planes \cite{Teng2022,Yi2024}. 
Within each FM layer, magnetic order breaks the degeneracy of electronic bands, splitting the spin-majority and spin-minority electron bands. 
One proposed origin of the 2$\times$2 CDW order in FeGe is the nesting of van Hove singularities (vHSs) near $E\rm_F$ in the spin-polarized electronic bands below $T\rm_{N}$ \cite{Teng2022}. 
In this scenario, the variation of the energy gap with temperature for vHSs at the M point urgently needs to be detected.

Distinct from the electronically driven mechanism from vHSs nesting, another proposed origin of the CDW order is dominated by Ge $z$-axis dimerization without a Kohn anomaly in electron-phonon coupling but with spin-charge-lattice coupling \cite{Miao2023,Teng2024,Yi2024,Wu2024,Chen2023}. 
The chirality of charge order anisotropy was correlated with antiferromagnetism, indicating a strong magnetic coupling of the charge order with magnetism \cite{Yin2022,Teng2023}. The CDW can also be tuned \cite{Chen2023,Shi2023}.

Theories propose that the charge order driven by kagome vHSs can induce a nontrivial Berry phase and various Berry phase-related quantum effects, including orbital currents/magnetism and bulk-boundary correspondence \cite{Zhou2022,RMP20102,Yin2022}. 
Meanwhile, STM studies of the Fe$_3$Ge kagome layer in FeGe present three pairs of 2$\times$2 vector peaks with different intensities, leading to orbital currents and weak orbital magnetization with a nontrivial Berry phase, resembling the chiral charge order reported in AV$_3$Sb$_5$ \cite{Yin2022}.

Topological charge order in FeGe can also feature bulk-boundary correspondence with edge states inside the bulk gap, confirming the nontrivial bulk gap with gapless boundary states \cite{Yin2022}. Thus, experimental observations of the kagome material FeGe provide a rare type of AFM-CDW phase to explore the origin of the CDW order and the relationships between lattice, spin, orbital, topology, and electronic correlation.
	
	In paramagnetic CoSn of the $TX$ family, the characteristic clean kagome bands, DP and FB, with degeneracy near the Fermi level, typically emerge in materials due to the relatively isolated paramagnetic kagome lattices, contrasting with the kagome magnet $T_3X$, as illustrated in Fig.~\ref{F3}(d). 
	The ARPES study reported that in CoSn, there are three sets of kagome bands in the 3$d$ orbitals of Co, $d_{xy}$/$d_{x^2-y^2}$, $d_{xz}$/$d_{yz}$, and $d_{z^2}$, while the different orbital components of the Dirac cone and FB exhibit distinct orbital-selective responses in polarized measurements due to the matrix element effect. Considering SOC, the DPs of various orbital components will result in distinct gap sizes, with significantly stronger SOC in the in-plane orbitals compared to the out-of-plane ones. This discrepancy is also a consequence of orbital selection \cite{Liu2020}. 
	
	The photon-energy-dependent ARPES measurement also presents the FB near $E_\mathrm{F}$ across the whole BZ with negligible dispersion along the out-of-plane momentum $k_z$ \cite{Kang2020}. To assess the degree of correlation in the CoSn system, a renormalization factor of $\sim 1.5$ was applied, which matched well with the main ARPES features in Fig.~\ref{F3}(f), indicating intermediate electron correlation effects in CoSn \cite{Liu2020}.

	Calculations of spin Hall conductivity in CoSn show a nontrivial topology with a nonzero $\mathbb{Z}_2$ invariant, which is concentrated near the spin-orbit coupling gap between the FB and quadratic band at $\Gamma$ and the massive Dirac points at $K$ \cite{Kang2020}.	
	STM reported a pronounced double kink feature in the kagome layer of CoSn, different from the dispersion calculated by first principles. This feature is identified as a fingerprint of Bosonic mode, proving the existence of many-body interactions between the dispersive band and the flat-band phonon \cite{Yin2020}.
	
	In transport measurements, the resistivity within the kagome plane is more than one order of magnitude larger than the $z$-axis resistivity, which is in sharp contrast with conventional 2D layered materials. Moreover, the magnetic susceptibility under an out-of-plane magnetic field is found to be much smaller compared to the in-plane case \cite{Huang2022}. These anomalous and giant anisotropies can be reasonably attributed to the unique properties of flat-band electrons with the closest $E_\mathrm{F}$ along the $L$--$H$--$A$ line and could contribute to the negative orbital diamagnetism illustrated in Fig.~\ref{F3}(h) \cite{Huang2022}.
	
	Considering the proximity of the $E_\mathrm{F}$ to the FB at $H$ in CoSn, once the FB is precisely tuned to $E_\mathrm{F}$, the significant DOS within the FB will precipitate the emergence of correlated phases. Paramagnetic CoSn, when partially substituting tin with nonmagnetic indium (CoSn$_{1-x}$In$_x$), tunes the Fermi energy into the FB region and induces magnetic instability, leading to an itinerant A-type AFM order with small moments lying within the kagome planes \cite{Sales2022}. 
	The observation of a topological FB in the kagome lattice CoSn opens up a new avenue to study correlation-driven emergent electronic phenomena against the backdrop of topological nontriviality.

	\section{Rich magnetism and topology in double-layered kagome magnet $AT_6X_6$} \label{AT166}
		
	Another kagome metal family, $AT_6X_6$ ($A$ = Li, Mg, Ca, or rare earth; $T$ = V, Cr, Mn, Fe, Co, Ni; $X$ = Sn, Ge), known as the `166' family, has attracted substantial attention due to its prevalent magnetic properties and large chemical diversity. Here, we mainly focus on two series: the V-166 ($A$V$_6$Sn$_6$) system and the Mn-166 ($A$Mn$_6$Sn$_6$) system. 
	
	The $AT_6Sn_6$ compound is composed of A-Sn3, T-Sn1, and Sn2 layers stacked along the $z$-axis, comprising two kagome layers per unit cell. The interval between kagome layers is greater than that in $T_3X$  family but slightly greater than in $TX$ family, suggesting a quasi-two-dimensional  characteristic and allowing a direct study of the electronic structure of the kagome layer in the 166 system. However, complex interlayer and intralayer interactions occur within the kagome lattices and sublattices of other elements, implying that the 166 system will manifest exotic quantum phases and could introduce tunable magnetism.
	
	In $A$V$_6$Sn$_6$, non-magnetic or weakly magnetic properties typically serve as a good platform for tuning nontrivial band topology associated with the non-magnetic transition metal V kagome, with anisotropic magnetic properties arising solely from the rare earth triangular lattice of the A-site. 
	The magnetic ordering temperature of $A$V$_6$Sn$_6$ is $T_\mathrm{N}$ = 4.8, 4.3, 2.9, and 2.3 K for $R$ = Gd, Tb, Dy, and Ho, respectively \cite{Lee2022,Pokharel2022,Ishikawa2021,Pokharel2022}. 
	The compounds YV$_6$Sn$_6$ and LuV$_6$Sn$_6$ exhibit typical characteristics of paramagnetic metals, and no magnetic ordering is observed down to low temperatures for ErV$_6$Sn$_6$, TmV$_6$Sn$_6$, and SmV$_6$Sn$_6$ \cite{Pokharel2021,Lee2022,Huang2023}.
	
	Meanwhile, in the $A$V$_6$Sn$_6$ system, the Fermi level is mainly contributed by V-3$d$ electrons, typically exhibiting FB, vHS, and DP near $E_\mathrm{F}$ in calculations or experiments, e.g. in $A$ = Y, Gd, Ho, Sm \cite{Ishikawa2021,Pokharel2021,Peng2021,Hu2022,Huang2023}. 
	Recently, theoretical predictions suggest the existence of $\mathbb{Z}_2$ topological nontrivial states with Dirac surface states in most paramagnetic $A$V$_6$Sn$_6$, and this has been partially confirmed in experiments, such as in GdV$_6$Sn$_6$, YV$_6$Sn$_6$, HoV$_6$Sn$_6$, and ScV$_6$Sn$_6$ systems \cite{Pokharel2021,Hu2022,Peng2021,Hu2024}. 
	The $A$V$_6$Sn$_6$ system also offers an intriguing platform for manipulating Dirac surface states and vHSs through potassium dosing, which is essential for tailoring topological characteristics and investigating potential spintronic applications \cite{Hu2022}.

	A newly discovered kagome metal, ScV$_6$Sn$_6$, reveals a first-order phase transition around T= 92 K and exhibits $\sqrt{3} \times \sqrt{3} \times 3$ CDW orders \cite{Arachchige2022}. Additionally, a theoretical study proposed a large lattice instability with abundant CDW orders in ScV$_6$Sn$_6$ due to the small atomic radius of Sc \cite{Tan2023}. X-ray experiments have detected a dynamic short-range $\sqrt{3} \times \sqrt{3} \times 2$ CDW order above 92 K, which competes with the $\sqrt{3} \times \sqrt{3} \times 3$ CDW order \cite{Cao2023}. It is further suggested that both physical and chemical pressure can quench this effect \cite{Gu2023}.
	
	In the ScV$_6$Sn$_6$, the electronic structure shows the existence of multiple saddle points near the $E_\mathrm{F}$ contributed by the V-kagome layer, as well as Dirac points at the $K$ point. A large electron pocket centered at the $\Gamma$ point is mainly contributed by Sn atoms near the kagome layer \cite{Hu2024,Lee2024}. 
	Tracking the temperature evolution of the band structure across the charge order transition reveals that the Sn band at $\Gamma$ develops a charge order gap and reconstructs the Fermi surface, while the V kagome bands remain unaltered, elucidating the significant role of Sc atoms in the formation of the CDW \cite{Lee2024}.
	
	The incorporation of Cr-doped ScV$_6$Sn$_6$, Sc(V$_{1-x}$Cr$_x$)$_6$Sn$_6$, results in the Fermi level shifting, causing the vHSs to move away from the $E_\mathrm{F}$. 
	 Meanwhile, the CDW order remains stable over a wide doping range, affirming the marginal role of vHSs at the $M$ point \cite{Lee2024}. 
	Phonon calculations for ScV$_6$Sn$_6$ display a continuum of unstable phonon modes centered at $H$.
	 Associated with structural distortions involving the planar Sn and Sc sites,  the contribution from the V kagome layer is negligible \cite{Lee2024,Hu2024}.
	
	Raman measurements reveal a two-phonon mode in the normal state and new emergent phonon modes with high frequency in the CDW phase, indicating a strong electron-phonon coupling \cite{Hu2024,Gu2023}. 
	The $\sqrt{3} \times \sqrt{3} \times 3$ CDW order in ScV$_6$Sn$_6$ is also observed to be sensitive to both physical and chemical pressure, disappearing around P = 2.4 GPa without the concurrent emergence of superconductivity \cite{Zhang2022}.
	
	Recently, a newly discovered kagome metal, YbV$_6$Sn$_6$, shows typical heavy-fermion properties due to the Kondo effect on the Kramers doublet of Yb$^{3+}$ ions above T = 2 K, with a remarkable magnetic ordering occurring at $T_\mathrm{N}$ = 0.40 K. 
	This provides a rare example of heavy-fermion kagome compounds hosting diverse FBs derived from the 4$f$-orbital of the Yb Kondo lattice or the 3$d$-orbital of the V kagome lattice \cite{Guo2023}. 
	To confirm the FB characteristics near the $E_\mathrm{F}$ in the YbV$_6$Sn$_6$ system, future electronic structure measurements are urgently needed.
	
\begin{figure*}[htb]
		\begin{center}
			\includegraphics[width=1\textwidth]{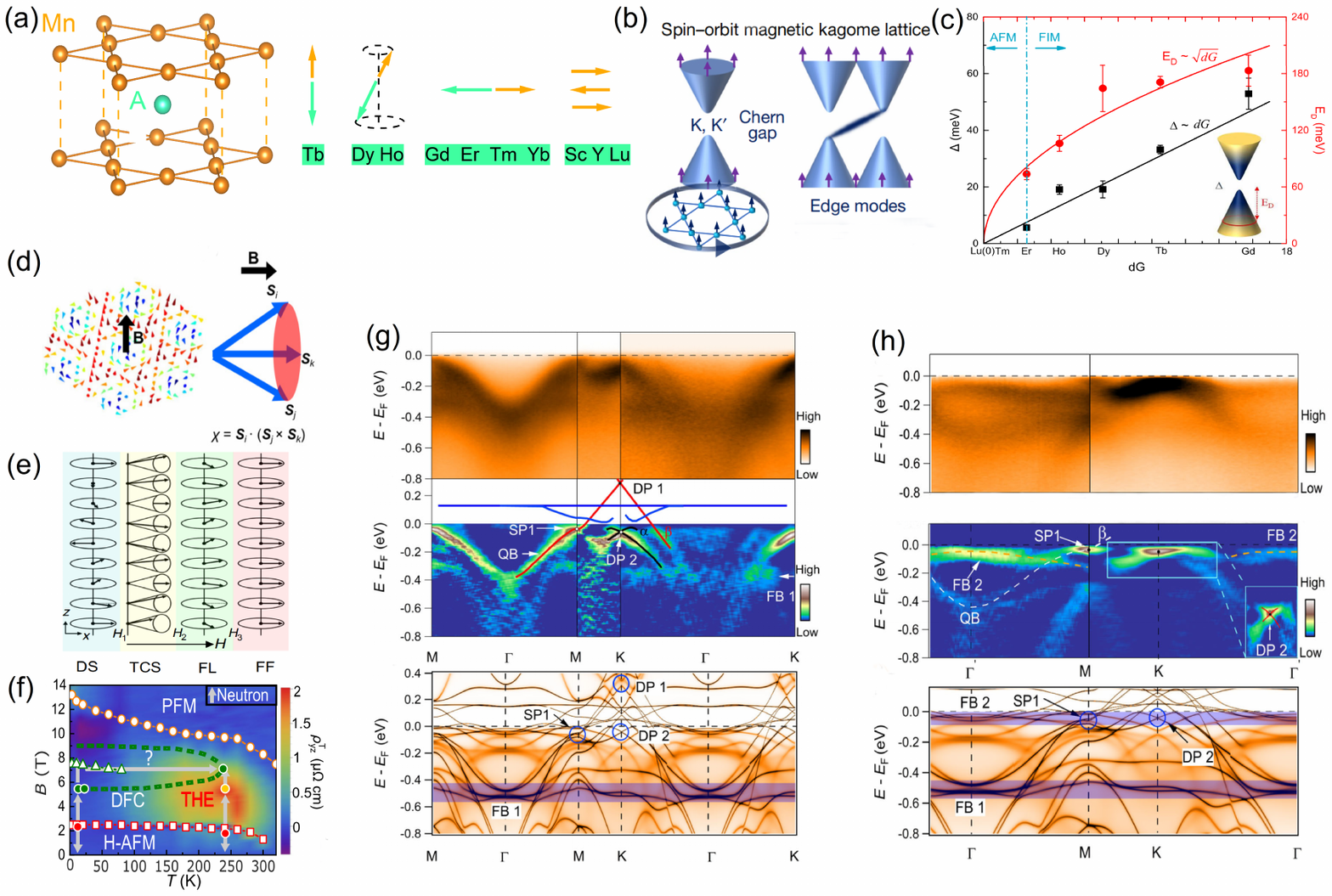}
			\caption{ \textbf{Crystal structure and electronic structure of $A$Mn$_6$Sn$_6$.}			
				(a) Two kagome layers are connected by an A atom layer in $A$Mn$_6$Sn$_6$, where the magnetic states of the A atom and Mn differ in various $A$Mn$_6$Sn$_6$ compounds.
				(b) Band structure of spin–orbit-coupled kagome magnet TbMn$_6$Sn$_6$, where the spin-polarized Dirac bands open a Chern gap with an edge mode arising in the gap (adapted from Ref.\cite{YGX2020}).
				(c) The systematic evolution of Dirac cone energy $E_{\mathrm{D}}$ and gap size $\Delta$ follows $\sqrt{dG}$ and $dG$, respectively (adapted from Ref.\cite{MWL2021}).
				(d) A schematic of a nanoscale skyrmion lattice with a field applied perpendicular to the kagome lattice plane, showing the nonzero scalar spin chirality $\chi$ with three non-coplanar spins $S_i$, $S_j$, and $S_k$ (adapted from Ref.\cite{WQ2021}).
				(e) Schematic of different field-induced magnetic structures in YMn$_6$Sn$_6$ (adapted from Ref.\cite{Ghimire2020}).
				(f) Phase diagram of YMn$_6$Sn$_6$ for $B\parallel[100]$ with a contour plot of $\rho_{\mathrm{yz}}^T(T,B)$. $M(B)$ data points are in white with colored outlines, and neutron data points are in color with white outlines. The H-AFM, DFC, and PFM indicate the helical AFM structure, double-fan phase with a z-axis component, and polarized ferromagnetic structure, respectively (adapted from Ref.\cite{WQ2021}).
				(g) Illustration of the photoemission intensity plots, second derivative plots, and DFT + DMFT calculated ARPES in the FM state with SOC of YMn$_6$Sn$_6$ along $\Gamma$--$M$--$K$--$\Gamma$ in the $k_z \sim 0$ plane, respectively.
				(h) Same as (g), but in the 2nd BZ \cite{Li2021}.
			}\label{F4}
		\end{center}
	\end{figure*}

	Contrasting with the V-166 based kagome lattice materials with a non-magnetic V kagome layer, the magnetic properties of the  magnetic  $A$Mn$_6$Sn$_6$ family change dramatically and behave sensitively to the rare-earth metal $A$ sublattice magnetic coupling with each FM kagome layer. In the Mn-166 compound, complex magnetic textures originating from the A-site magnetism and Mn-based FM kagome layers generally appear, exhibiting diverse long-range magnetic orders including ferromagnetism ($A$ = Li, Mg, Ca, Nd, Sm, Yb) \cite{Mazet2006,Chen2021,Ma2021,Mazet2002,Song2024,LLF2024}, ferrimagnetism ($A$ = Tb, Dy, Ho, Er, Tm, Gd) \cite{Gao2021,Venturini1991,Xu2022,Zeng2022,Kabir2022,Dhakal2021,Suga2006,Riberolles2024,Wang2022,Gorbunov2012}, and antiferromagnetism or helical antiferromagnetism ($A$ = Sc, Y, Lu) \cite{Chafik1991,ZH2022,WQ2021,Ghimire2020,MWL2021}. 
	Figure~\ref{F4}(a) displays a sketch of the bilayer kagome structure of $A$Mn$_6$Sn$_6$ and the magnetic state of the $A$ atom (green arrow) and Mn (yellow arrow) in different $A$Mn$_6$Sn$_6$ compounds. 
	
	Recently,  novel electromagnetic properties have been revealed in the $A$Mn$_6$Sn$_6$ system. For instance, most $A$Mn$_6$Sn$_6$ kagome magnets exhibit an intrinsic anomalous Hall effect derived from the nontrivial Berry curvature, except for TmMn$_6$Sn$_6$ and LuMn$_6$Sn$_6$ where nontrivial massive Dirac fermions generally exist in those kagome lattices \cite{Asaba2020,WQ2021,MWL2021,Gao2021,Dhakal2021,Chen2021,Ma2021,Zeng2022,ZH2022,Lv2023,LLF2024}. 
	The remarkable THE has also been detected in some $A$Mn$_6$Sn$_6$ kagome magnets, originating from the field-induced non-zero chiral spin texture \cite{Ghimire2020,WQ2021,Dhakal2021,ZH2022,Liu2023}, as shown in Fig.~\ref{F4}(d).
	
	The linear band crossing at $K$ and the quadratic band touching degeneracy at $\Gamma$, considering SOC, can be a singular source of Berry curvature and nontrivial topology \cite{Guo2009,Tang2011,Xu2015}. 
	In particular, a near-ideal quantum-limit magnet with Chern-gapped massive Dirac fermions with out-of-plane ferrimagnetic (FIM) in TbMn$_6$Sn$_6$ has been identified using STM and quantum oscillations, which exhibits a non-trivial in-gap edge state, as shown in Fig.~\ref{F4}(b) \cite{YGX2020,MWL2021}. 
	The Chern Dirac fermion generally exists in the Mn-based kagome lattices with a net magnetic moment in $A$Mn$_6$Sn$_6$ for $A$ = Gd - Er, whose Berry curvature field generates a large intrinsic anomalous Hall effect \cite{MWL2021}. 
	
	Figure~\ref{F4}(c) shows the systematic evolution of the derived Dirac cone energy $E_\mathrm{D}$ and gap size $\Delta$ following $\sqrt{dG}$ and $dG$, respectively, where the de Gennes factor $dG$ = $(g_J-1)^2J(J+1)$, $g_J$ is the Lande factor, and $J$ is the total angular momentum of the $R^{3+}$ ion Hund’s rule ground state \cite{MWL2021}.

	Among the rare-earth metal alloys $A$Mn$_6$Sn$_6$, the YMn$_6$Sn$_6$ stands out due to its complex magnetic structure, which features a double magnetic sublattice. 
	It undergoes a paramagnetism–antiferromagnetism phase transition at the N\'eel temperature T$\rm_N$ = 359K, transitioning from AFM order along the z-axis to a helical AFM order with in-plane FM order below T = 326 K \cite{Ghimire2020,WQ2021}.
	
	In YMn$_6$Sn$_6$, with increasing magnetic field (B$\|$[100]), the magnetic state transitions from a distorted spiral (DS) state to a transverse conical spiral (TCS) phase, a fan-like state (a quadrupled structure along the $z$-axis with spins deviating from the in-plane magnetic field direction), and finally to a forced ferromagnetic (FF) state, as shown in Fig.\ref{F4}(e) \cite{Ghimire2020}. 
	Figure~\ref{F4}(f) presents a complex phase diagram of YMn$_6$Sn$_6$ for B$\|$[100], illustrating the helical AFM (H-AFM), double-fan phase with a $z$-axis (DFC), and polarized FM (PFM) states corresponding to the DS state, TCS state, and FF state in Fig.~\ref{F4}(e), respectively \cite{WQ2021}. The area denoted as "?" in Fig.\ref{F4}(f) was revealed to be a FL state by Siegfried {\it{et al}}. \cite{Siegfried2022}.
	
	Particularly, the THE in YMn$_6$Sn$_6$ is proposed to originate from dynamic chiral fluctuation-driven non-zero spin chirality in the DFC phase, rather than from a field-driven skyrmion lattice \cite{WQ2021}. 
	ARPES measurements reveal that the band structure of magnetic kagome YMn$_6$Sn$_6$ exhibits the complete characteristics of the kagome lattice near $E\rm{_F}$, 
	including DP, vHS, and FB. Spin-polarized DFT + DMFT calculations indicate that the flat band (FB2, Fig.~\ref{F4}(h)) with a high density of states around $E\rm{_F}$ is from the majority-spin state, while the Dirac point (DP2) is from the minority-spin state, indicating its single-spin degenerate origin \cite{Li2021}.
	
	The massive Dirac point (DP2) and flat band (FB2) near $E\rm{_F}$, arising from the spin-polarized band with SOC, exhibit intrinsic Berry curvature that may explain the anomalous Hall effect observed in transport measurements \cite{Li2021}. 
	Theoretically, the non-trivial Berry curvature associated with spin-polarized massive Dirac fermions typically generates an orbital magnetic moment. 
	STM measurements confirm the presence of large orbital magnetic moments and reveal a momentum-dependent effective $g$-factor associated with a nontrivial massive Dirac band in a kagome magnet \cite{Li2022}. 
	Since the magnetic order at the surface of YMn$_6$Sn$_6$ breaks time-reversal symmetry, the gapped Dirac band should carry a non-zero Chern number, indicating that YMn$_6$Sn$_6$ is a topological kagome lattice Chern magnet \cite{Li2022}.
	
	Although $A$Mn$_6$Sn$_6$ compounds exhibit various complex magnetic behaviors, such as z-axis FIM in TbMn$_6$Sn$_6$, canted z-axis FIM in DyMn$_6$Sn$_6$, in-plane FIM in GdMn$_6$Sn$_6$, and in-plane helical antiferromagnetism in YMn$_6$Sn$_6$, they all display robust band structures
	 (DP and FB near $E\rm{_F}$), 
	suggesting that magnetic order plays a secondary role in influencing the band structure \cite{Li2021,Gu2022,LZH2021}.
	
	Moreover, unlike most $A$Mn$_6$Sn$_6$ compounds which exhibit intrinsic AHE with a net magnetic moment, some compounds like LuMn$_6$Sn$_6$ without intrinsic AHE allow trivial Dirac dispersion. Future experiments probing the electronic structure of trivial 166-Mn kagome magnets may provide further insights necessary to disentangle the connection between magnetism and topological properties.

	\section{Rich correlations and topology in kagome metal $R$V$_3$Sb$_5$} \label{Origins}
	
	$R$V$_3$Sb$_5$ ($R$ = K, Rb, Cs) is another representative kagome family, which has aroused much concern due to rich physical states, such as CDW, superconductivity, nematic and stripe order, PDW phase, $\mathbb{Z}_2$ topological state, large AHE, and chirality under magnetic field, etc. These complex, symmetry-breaking ordered states compete or intertwine, echoing high-temperature superconductors, yet their correlation remains enigmatic. The early works on $R$V$_3$Sb$_5$ were introduced in some reviews \cite{Neupert2022,Wang2023}.
	
\begin{figure*}[!htbp] 
		\begin{center}
			\includegraphics[width=1.0 \hsize]{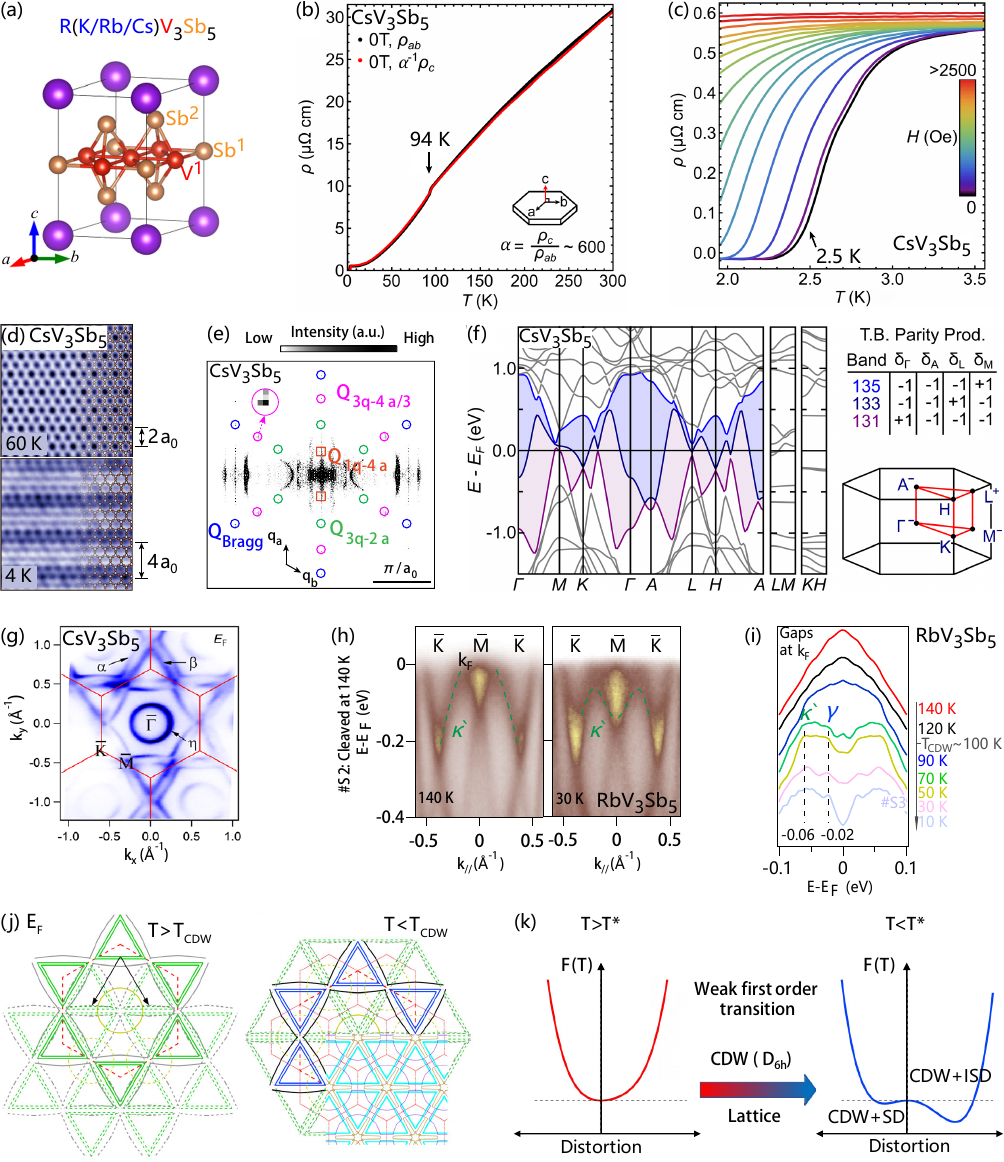} 
			\caption{
				(a) The structure of $R$V$_3$Sb$_5$ ($R$ = K/Rb/Cs) (adapted from Ref.\cite{Ortiz_2019}).
				(b) Full temperature ranges, and 
				(c) Field-dependent measurements at low temperatures 
				for the electrical resistivity of CsV$_3$Sb$_5$ (adapted from Ref.\cite{Ortiz_202102}).
				(d) Atomically resolved STM topographs, taken over the Sb-terminated surface (adapted from Ref.\cite{2021Cascade-ZhaoHe}).
				(e) Fourier transform of d$I/$d$V$ map of CsV$_3$Sb$_5$ (adapted from Ref.\cite{CH_2021}).
				(f) Calculated band structure of CsV$_3$Sb$_5$. Shaded areas show continuous direct gaps (adapted from Ref.\cite{Ortiz_202102}).
				(g) Fermi surface of ARPES intensity map of CsV$_3$Sb$_5$ \cite{2022Charge-LouRui}.
				(h) ARPES intensity plots along the $\overline{\mathrm{K}}$--$\overline{\mathrm{M}}$ of RbV$_3$Sb$_5$, taken at 140 and 30 K \cite{LZH_2021}.
				(i) The symmetrized EDCs at the $k_{\mathrm{F}}$ in (d) \cite{LZH_2021}.
				(j) Sketches of the experimental Fermi surfaces (solid curves) with the folded ones (dashed curves) by the $2 \times 2$ CDW vector, which are above and below $T_{\mathrm{CDW}}$ \cite{2022Charge-LouRui}.
				(k) CDW lattice coupling induced a weak first-order phase transition (adapted from Ref.\cite{MH_2021}).
			} 
			\label{F5}
		\end{center}
	\end{figure*}

	As shown in Fig.~\ref{F5}(a), the crystal structure of $R$V$_3$Sb$_5$ crystallizes in a hexagonal structure with the $P6/mmm$ (No. 191) space group \cite{Ortiz_2019}. There are V-Sb slabs consisting of V kagome nets and interspersing Sb atoms, which are separated by $R$ ions along the $z$ axis. There are two kinds of Sb sites: the Sb$^1$ site at the centers of V kagome hexagons, and the Sb$^2$ site below and above the centers of V triangles forming graphenelike hexagon layers.
	
	In $R$V$_3$Sb$_5$, there are not only CDW but also superconductivity \cite{Du2021, CKY2021, LHX2021, Ortiz_2021, Ortiz_202102, CPL_2021}. For example, the full temperature ranges for the electrical resistivity of CsV$_3$Sb$_5$ in Fig.~\ref{F5}(b) indicate the presence of an anomaly at T = $94$ K ($T_{\mathrm{CDW}}$) suspected to be an electronic instability (e.g., charge ordering), and field-dependent measurements at low temperatures in Fig.~\ref{F5}(c) show that superconductivity occurs at approximately $T_{\mathrm{c}} = 2.5$ K \cite{Ortiz_202102}. Unlike well-known CDW materials \cite{PhysRevLett.51.138, PhysRevX.8.011008, PhysRevLett.107.107403, PhysRevLett.107.266401, science.aam6432, PhysRevLett.35.120, PhysRevLett.102.086402}, CDW in $R$V$_3$Sb$_5$ fails to induce acoustic phonon anomalies near the CDW wave vector, indicating a strong commensurability effect \cite{LHX2021, MH_2021}.

$R$V$_3$Sb$_5$ shows abundant structures of the CDW phase, for example, unidirectional CDW \cite{CH_2021, LZW_2021, 2021Cascade-ZhaoHe}, a three-dimensional CDW with a $2 \times 2 \times 2$ superstructure \cite{Ortiz_202103} or a $2 \times 2 \times 4$ superstructure \cite{LZW_2021}, a chiral CDW \cite{Shumiya_2021, Mielke_2022}, and a CDW that breaks the six-fold rotational symmetry of the kagome lattice \cite{2021Cascade-ZhaoHe, LH_2022}. Correspondingly, two-fold rotational symmetry that persists into the superconducting phase was found in magnetoresistance measurements \cite{CH_2021, NSL_2021, XY_2021}. Starting at a high temperature, temperature-dependent STM of such a system is carried out \cite{2021Cascade-ZhaoHe}. As shown in the upper panel of Fig.~\ref{F5}(d), STM topographs at T = 60 K exhibit a $2 \times 2$ superstructure, which breaks the translational symmetry of the lattice. As the system is cooled down further, the electronic structure of CsV$_3$Sb$_5$ begins to display a pronounced unidirectional character. As shown in the lower panel of Fig.~\ref{F5}(d), another periodic modulation with a $4a_0$ wavelength at approximately T = 50 K propagates along only one lattice direction and emerges in STM topographs. At T = 300 mK, the magnitude of the drift-corrected, two-fold symmetrized Fourier transform in Fig.~\ref{F5}(e) reveals wave vectors $\boldsymbol{Q}_{3q - 2a} = \frac{1}{2} \boldsymbol{Q}_\mathrm{Bragg}$ for $2 \times 2$ CDW, $\boldsymbol{Q}_{1q - 4a} = \frac{1}{4} \boldsymbol{Q}_\mathrm{Bragg}$ for $4a_0$ unidirectional charge order, and $\boldsymbol{Q}_{3q - 4/3a}$ for new 3Q PDW modulations \cite{CH_2021}. $R$V$_3$Sb$_5$ also shows a magnetic response of the chiral charge order, suggesting a time-reversal symmetry breaking of the CDW order \cite{JYX_2021}.

	A rich interplay between CDW and superconductivity was observed in $R$V$_3$Sb$_5$ under external hydrostatic pressure \cite{Du2021, 2021Unusual-YuFH, CKY2021, 2021Highly-CPL, 2022Role-Alexander}, and CDW is monotonically suppressed with increasing pressure. Such competition between the CDW state and superconductivity is usual since the gap opening at the CDW state would dramatically reduce the density of states at Fermi surfaces, leading to the suppression of superconductivity within the Bardeen-Cooper-Schrieffer scenario. By comparing phase diagrams derived from uniaxial strain measurements and past hydrostatic pressure experiments, researchers discovered that the increase in $T_{\mathrm{c}}$ under pressure or tensile a-axis strain occurs because the competing CDW order is suppressed by changes in the $z$-axis lattice parameter. Both $T_{\mathrm{c}}$ and $T_{\mathrm{CDW}}$ are dominated by changes in the $z$-axis lattice parameter, regardless of whether they are promoted by hydrostatic pressure or uniaxial strain. Therefore, this comparison further suggests that the effect of the broken rotational symmetry induced by the uniaxial strain on the CDW and SC states is weak \cite{2021Revealing-QianTiema}.
	
	The entire $R$V$_3$Sb$_5$ system exhibits a uniform band structure, exemplified by the calculated band structure of CsV$_3$Sb$_5$, as depicted in Fig.~\ref{F5}(f). Dirac linear dispersion is found at the $K$ point, while a saddle point can be seen at the $M$ point. Competing density wave instabilities may also arise, and, in the present case, scattering wave vectors would connect an enhanced density of states at saddle points near the Fermi energy at the $M$ points. Nesting across a 2D Fermi surface with an underlying hexagonal motif is also thought to promote the formation of a superconducting state \cite{Ortiz_202102}. Theoretically, topologically nontrivial surface states close to $E_{\mathrm{F}}$ and the continuous direct gap throughout the Brillouin zone allow the identification of the normal state as a $\mathbb{Z}_2$ topological metal \cite{2015Z2PRB, 2017Z2NC}. Experimentally, the intercept on the Landau level index in the Landau level fan diagram of quantum transport experiments confirms the nontrivial topological property of CsV$_3$Sb$_5$ \cite{2021Quantum-FuYang}.

	From the perspective of band structure, two kinds of saddle points correspond to vHS (including m-type vHS and p-type vHS) have been observed by ARPES \cite{LZH_2021, SDW_2021, 2022Richnature-NC, 2022Twofold-NP, 2022Charge-LouRui}.
	High quality ARPES map of CsV$ _3 $Sb$ _5 $ in Fig.~\ref{F5}(g) \cite{2022Charge-LouRui} shows a high intensity around the $\overline{M} $ points, suggesting the singularities or the surface states at the proximity of $ E_{\mathrm{F}} $.
	Figure~\ref{F5}(h) shows the temperature evolution of the bands along the $\overline{K} $--$ \overline{M} $ direction. 
	The bands near the $\overline{M} $ point are remarkably renormalized by the CDW, while at T = 140 K, $ \kappa ^\prime $ band crosses $ E_{\mathrm{F}} $; With the decreasing temperature, $ \kappa ^\prime $ band shifts down.
	And at approximately T = 50--70 K, the $ \kappa ^\prime $ band centered at $\overline{M} $ is flattened and sinks below $ E_{\mathrm{F}} $, forming an M-shaped band with the tips of band (singularities) at about 60 meV below $ E_{\mathrm{F}} $.
	The symmetrized EDCs in Fig.~\ref{F5}(i) show that both $ \kappa ^\prime $ and $ \gamma $ bands along $  \overline{K} $--$\overline{M}$ further open the energy gap \cite{LZH_2021}.
	The electronic instability via Fermi surface nesting could play a role in determining these CDW-related features. 
	Above $ T_{\mathrm{CDW}} $, nearly perfect nestings can be seen in the left panel of Fig.~\ref{F5}(j). 
	Once upon entering the CDW phase, as shown in the right panel of Fig.~\ref{F5}(j), the interactions involved in nested bands and overlaps could result in gap opening and FS reconstructions in the CDW BZ. 
	The finite density of states at $ E_{\mathrm{F}} $ in the CDW phase is most likely in favor of the emergence of multiband superconductivity, particularly the enhanced density of states associated with the CDW-induced flat features just below $ E_{\mathrm{F}} $ \cite{2022Charge-LouRui}. 
	
	Based on the tight-binding model of the kagome lattice, $H_0=\sum_k c_k^{\dagger} H_k c_k$, there are possible charge-order states predicted for the $R$V$_3$Sb$_5$ family in Table~\ref{table_KagomeOrder} \cite{Feng2021}. The vector CDW can be obtained if the charge density for each sublattice is considered. Furthermore, the hopping bonds inside each unit cell are considered. If the hopping bond order parameters are real, one can get a charge bond order. If the hopping bond order parameters are imaginary, one can get a chiral flux phase, which breaks time-reversal symmetry. There are also Star of David (SD) and inverse Star of David (ISD) \cite{Wang2023}. As shown in the left panel of Fig.~\ref{F5}(k), DFT calculations show that the ideal kagome lattice is stable and corresponds to the free-energy minimum at high temperature. However, as shown in the right panel of Fig.~\ref{F5}(k), the lattice energy is asymmetric with respect to the lattice distortion at zero temperature: the SD is a local energy minimum, while the ISD is the global minimum. Since CDW always couples with lattice distortions through finite electron-phonon coupling, the asymmetric lattice-distortion free energy adds a cubic term in the CDW free energy through electron-phonon coupling near $T_{\mathrm{CDW}}$, which consequently drives the CDW transition to a weak first-order transition \cite{MH_2021}.
	
	Not long after the discovery of $R$V$_3$Sb$_5$, two new members of V-based kagome metals, CsV$_8$Sb$_{12}$ and CsV$_6$Sb$_6$, were reported \cite{YYX_2021, YQW_2021}. By inserting alternative building blocks or changing the stacking sequences, the initial $C_6$ symmetry in $R$V$_3$Sb$_5$ decreases to $C_2$ in $R$V$_8$Sb$_{12}$ and $C_3$ in $R$V$_6$Sb$_6$ \cite{YYX_2021}. 
	However, CsV$_8$Sb$_{12}$ and CsV$_6$Sb$_6$ are distinctly different from $R$V$_3$Sb$_5$ because the CDW and superconducting transitions are absent in both CsV$_8$Sb$_{12}$ and CsV$_6$Sb$_6$ down to about T = 2 K \cite{YYX_2021, YQW_2021}. There are two types of saddle points (vHSs) in CsV$_8$Sb$_{12}$ \cite{2023Mahuan-CsV8Sb12PRB}. 
	One vHS at $\overline{\mathrm{M}}$ originates from the kagome lattice and is located at high binding energy. Compared to CsV$_3$Sb$_5$ \cite{2022Charge-LouRui}, the saddle point of the V-based kagome in CsV$_8$Sb$_{12}$ is located further away from $E_{\mathrm{F}}$. The absence of vHSs near $E_{\mathrm{F}}$ could be the reason for the disappearance of CDW and superconductivity in CsV$_8$Sb$_{12}$. 
	Another vHS at $\overline{\Gamma}$ arises from the V$_2$Sb$_2$ layer and is situated near $E_{\mathrm{F}}$. The saddle point around $\overline{\Gamma}$ close to $E_{\mathrm{F}}$ enhances the density of states near $E_{\mathrm{F}}$. 
	However, being the only hot spot at the center of the BZ, the saddle point at $\overline{\Gamma}$ does not contribute to the intra-Fermi surface scattering. High-pressure and low-temperature measurements do not show any sign of superconductivity down to T= 0.3 K in CsV$_8$Sb$_{12}$ \cite{YQW_2021}. 
	However, superconductivity was found in $R$V$_6$Sb$_6$ upon applying quasi-hydrostatic pressures \cite{SMZ_2021}. 
	The high-pressure phase diagram suggests a complex interplay of the pressure-dependent density of states and structural instabilities. Such modification stems from a structural phase transition that occurs above P $\sim 20$ GPa, where the structure changes from rhombohedral to monoclinic. It is most likely that the monoclinic phase hosts superconductivity \cite{SMZ_2021}.

	\begin{table*}[!htbp] 
		\caption{ Possible charge-order states, based on the tight binding model of kagome $H_0=\sum_k c_k^{\dagger} H_k c_k$ \cite{Feng2021}. }
		\label{table_KagomeOrder} 
		\centering
		\begin{tabular}{llll}
			\toprule
			\midrule
			possible charge orders & \phantom{~~~~~~} Hamiltonian & \phantom{~~~~~~} order parameter &  \\
			\midrule
			vector CDW (vCDW) 
			& \phantom{~~~~~~} 
			$H_0 -\sum_R \boldsymbol{\Delta}_{\mathrm{vCDW}} \cdot \hat{\mathbf{n}}(\mathbf{R})$ 
			&
			\makecell[l]{ \phantom{~~~~~~} $\boldsymbol{\Delta}_{\mathrm{vCDW}}(\mathbf{R})=$ \\ \phantom{~~~~~~} $\lambda\left(\cos \left(\mathbf{Q}_a \cdot \mathbf{R}\right), \cos \left(\mathbf{Q}_b \cdot \mathbf{R}\right), \cos \left(\mathbf{Q}_c \cdot \mathbf{R}\right)\right)$ }
			& 
			\makecell[l]{ \phantom{~~~~~~} $\hat{\mathbf{n}}(\mathbf{R})=$ \\ \phantom{~~~~~~} $\left( \hat{n}_{\mathrm{A}}, \hat{n}_{\mathrm{B}}, \hat{n}_{\mathrm{C}} \right)$ } 
			\\
			\midrule
			charge bond order (CBO) 
			& \phantom{~~~~~~} 
			$H_0 -\sum_R \boldsymbol{\Delta}_{\mathrm{CBO}} \cdot \hat{\mathbf{n}}(\mathbf{R})$
			&  
			\makecell[l]{ \phantom{~~~~~~} $ \boldsymbol{\Delta}_{\mathrm{CBO}}(\mathbf{R}) = \boldsymbol{\Delta}_{\mathrm{vCDW}}(\mathbf{R}) $ } 
			&  
			\makecell[l]{ \phantom{~~~~~~} $\hat{\mathbf{O}}(\mathbf{R})=$ \\ \phantom{~~~~~~} $\left(c_{\mathrm{A}}^{\dagger} c_{\mathrm{B}}, c_{\mathrm{B}}^{\dagger} c_{\mathrm{C}}, c_{\mathrm{C}}^{\dagger} c_{\mathrm{A}}\right)$  } 
			\\
			\midrule
			chiral flux phase (CFP) 
			& \phantom{~~~~~~} 
			$H_0 -\sum_R \boldsymbol{\Delta}_{\mathrm{CFP}} \cdot \hat{\mathbf{n}}(\mathbf{R})$ 
			&  
			\makecell[l]{ \phantom{~~~~~~} $ \boldsymbol{\Delta}_{\mathrm{CFP}}(\mathbf{R}) = i \boldsymbol{\Delta}_{\mathrm{vCDW}}(\mathbf{R}) $ } 
			&  
			\makecell[l]{ \phantom{~~~~~~} $\hat{\mathbf{O}}(\mathbf{R})=$ \\ \phantom{~~~~~~} $\left(c_{\mathrm{A}}^{\dagger} c_{\mathrm{B}}, c_{\mathrm{B}}^{\dagger} c_{\mathrm{C}}, c_{\mathrm{C}}^{\dagger} c_{\mathrm{A}}\right)$  } 
			\\
			\midrule
			\bottomrule
		\end{tabular}
	\end{table*}
	
	Besides the aforementioned kagome systems, the "322" kagome family including  Co$_3$Sn$_2$S$_2$, Ni$_3$In$_2$S$_2$, and Ni$_3$In$_2$Se$_2$, is also attracting considerable attention. The kagome ferromagnet Co$_3$Sn$_2$S$_2$ with large intrinsic AHE is believed to host magnetic Weyl semimetal states\cite{LEK2018,Wang2018}. 
	Given a well-defined electronic band structure with reduced correlation effects, such magnetic Weyl semimetal states in Co$_3$Sn$_2$S$_2$ have been demonstrated to host superior electronic coherence compared to Mn$_3$Sn \cite{Kuroda2017,LDF2019}.	
    Direct experimental observations have confirmed the presence of both bulk Weyl fermions and surface Fermi arcs in Co$_3$Sn$_2$S$_2$, validating the existence of magnetic Weyl semimetal states \cite{LDF2019,Belopolski2021}. Additionally, 
    the optimal 2D high-temperature quantum anomalous Hall states (Chern insulator states) with 1D chiral edge states are expected to be realized in monolayer Co$_3$Sn$_2$S$_2$, which, due to its layered structure, can be achieved through mechanical exfoliation of the bulk crystals or the film growth\cite{Muechler2020,Howard2021}. 
    This provides previous opportunities for exploring more exotic topological phenomena in the layered ferromagnetic kagome materials. 
    For the kagome metals Ni$_3$In$_2$S$_2$ and Ni$_3$In$_2$Se$_2$, the high carrier mobility and substantial unsaturated magnetoresistance have been observed in the transport measurements, which can be attributed to the underlying Dirac nodal line semimetal states, as suggested by both the experiments and theory \cite{ZTT2022,Fang2023,Pradhan2024}.

	\section{{Conclusion and outlook}}

	In this topical review, we present an overview of recent studies on kagome materials with different kinds of structures, including $T_3X$, $TX$, $AT_6X_6$, and $RT_3X_5$ families. In kagome metals, the introduction of atomic layers provides a new pathway for tuning the system's magnetism and correlations, leading to rich long-range magnetic orders, nontrivial topological properties, and diverse correlation-related states such as CDW, superconductivity, fractional quantum Hall effect, intrinsic topological superconductivity, and so on. One may also expect the further manipulation of these states by external factors, such as chemical or electrical doping, strain, and pressure, which could lead to more exotic quantum phenomena.
	
	One can see that the kagome materials involved in our review exhibit at least one of the kagome band characteristics (Dirac cone, FB, and vHS), which are derived from the frustrated lattice geometry of the kagome lattice. Combining this unique lattice geometry with electron correlations and magnetism, novel quantum states can be induced. For example, the Dirac cone together with the $z$-axis net magnetic moment results in the Chern magnet; 1/3 or 2/3 filling of the FB can give rise to a high-temperature fractional quantum Hall state; the nesting of vHSs near $E_{\mathrm{F}}$ could trigger CDW instabilities. These studies pave the way for exploring novel electronic and magnetic behaviors in condensed matter physics. The intricate interplay between geometry and electron interactions in kagome lattices offers a rich playground for future theoretical and experimental investigations.
	
	Looking ahead, further studies can delve deeper into the understanding and manipulation of these emergent phenomena towards potential applications in quantum computing, spintronics, and beyond. Further efforts to synthesize and characterize new kagome materials are crucial not only for advancing the knowledge of the community but also for realizing the technological potentials of kagome materials. Moreover, the exploration of ideal 2D kagome materials through mechanical exfoliation, MBE growth, and intercalation will probably realize even more pristine FBs and cleaner 2D Dirac fermions. In this context, it might also be reasonable to expect the construction of kagome-based moiré superlattices in the near future.

	\begin{acknowledgments}
		This work was supported by the National Natural Science Foundation of China (Grants No. 12204536), the Fundamental Research Funds for the Central Universities, and the Research Funds of People's Public Security University of China (PPSUC) (2023JKF02ZK09).
		M.L., and H.M. contributed equally to this work.
	\end{acknowledgments}
	\bibliographystyle{iopart-num.bst}
	\bibliography{ref.bib} 

\end{document}